\documentclass{appolb}
\usepackage{epsfig}

\usepackage{amsmath}

\usepackage{graphicx}  
\usepackage{subfigure}

\usepackage{hyperref} 

\begin{document}
\title{ Mass spectrum and decays of the first radial excitation axial vector meson nonet 
}
\author{Xue-Chao Feng $^{1}$\thanks{fxchao@zzuli.edu.cn}, Ke-Wei Wei $^{2}$\thanks{weikw@ihep.ac.cn}
\address{$^{1}$ Department of Technology and Physics, Zhengzhou University of Light Industry, 450002 Zhengzhou, China
\\ $^{2}$ School of Science, Henan University of Engineering, 451191 Zhengzhou, China } }

\maketitle
\begin{abstract}
In this work, the mass spectrum of the first radial excitation axial vector meson nonet is considered in the framework of the nonrelativistic constituent quark model and Regge phenomenology. After that, we investigate the strong decay characteristics of these states within the $^{3}P_{0}$ model. The results are compared to values from other phenomenological models, and they may be beneficial in the prospective search for $2^{3}P_{1}$ meson nonets.

\end{abstract}
\PACS{11.55.Jy, 12.40.Yx, 14.40.Be}

\section{Introduction}

In the preceding decades, with the advancement of experiments, more and more light flavor mesons have been found. During this time period, both measurements and the theoretical interpretation of the results have made significant strides. In the most recent version of the Particle Data Group, light flavor mesons, especially the ground state light mesons, have been explicitly allotted \cite{ParticleDataGroup:2022pth}. However, we also note that the assignment of radial excited states is not optimistic. On the one hand, several candidates for the radial excited states have not yet been observed experimentally. On the other hand, there are several candidates for some radial excited states that need to be screened further through theory and experiment. There is still a significant amount of work that has to be done, both theoretically and practically, in order to finish and identify the meson spectrum.

 In this work, we concentrate on the ground and the first radially excited axial vector meson state. In Table 1, we present the experimental information of states with $J^{PC}=1^{++}$ quantum numbers in the Particle Data Group (PDG) \cite{ParticleDataGroup:2022pth}.
\begin{center}
\indent\\ \footnotesize Table 1. Masses and decay widths of $J^{PC}=1^{++}$ meson states (in units of MeV).  $^{\dag}$The states listed as  ``further states'' in the PDG \cite{ParticleDataGroup:2022pth}. $^{\ddag}$ The isodoublet of $1^{3}P_{1}$ meson nonet $K_{1A}$ is mixture of $K_{1}(1270)$ and $K_{1}(1400)$.
\label{Tab:t1}
\begin{tabular}{llllllllll}
\\ \hline\hline
  $I^{G}(J^{PC})$  & State ($1^{3}P_{1}$)       &Mass               & Width
\\
\hline
 $1^{+}(1^{++})$   &$a_{1}(1260)$             &  $1230\pm 40$             & $420\pm 35$          &
\\
                   &$a_{1}(1640)$             &  $1655\pm 16$             & $254\pm 40$          &
\\
                   &$a_{1}(1930)^{\dag}$      &  $1930^{+30}_{-70}$       & $155\pm 45$          &
\\
                   &$a_{1}(2095)^{\dag}$      &  $2096\pm 17 \pm 121$      & $451\pm 41 \pm 81$           &
\\
                   &$a_{1}(2270)^{\dag}$      &  $2270^{+55}_{-40}$       & $305^{+70}_{-40}$       &
\\ \hline
 $0^{+}(1^{++})$   &$f_{1}(1285)$            &  $1281.9\pm 0.5 $             & $22.7\pm 1.1$             &
\\
                   &$f_{1}(1420)$            &  $1426.3\pm 0.9 $             & $54.5\pm 2.6$             &
\\
                   &$f_{1}(1510)$            &  $1518\pm 5 $             &  $73\pm 25$             &
\\
                   &$f_{1}(1970)^{\dag}$     &  $1971\pm 15 $             & $240\pm 45$             &
\\
                   &$f_{1}(2310)^{\dag}$     &  $2310\pm 60 $             & $255\pm 70$             &
\\ \hline
$\frac{1}{2}^{+}(1^{++})$
                   &$K_{1}(1270)^{\pm\ddag}$    &  $1253\pm 7 $             & $90\pm 20$             &
\\
                   &$K_{1}(1400)^{\pm\ddag}$    &  $1403\pm 7 $             & $194\pm 17$             &
\\
\hline

\end{tabular}
\end{center}

For the ground axial vector meson nonet, the $a_{1}(1260)$, $f_{1}(1285)$ and $f_{1}(1420)$ are well established. The isodoublet state $K_{1A}$ is the mixture of $K_{1}(1270)$ and $K_{1}(1400)$, and the mixing angle has been estimated by different approaches \cite{Li:2006we,Cheng:2011pb,Divotgey:2013jba,Zhang:2017cbi}. On the whole, the assignment for the ground axial vector meson nonet remains plausible in the light of the available experiment data. However, under the existing experimental data, the assignment for the first radially excited state of axial vector meson nonet has encountered great difficulties. On the one hand, the $a_{1}(1640)$, $a_{1}(1930)$, $a_{1}(2095)$ and $a_{1}(2270)$ with $J^{PC}=1^{++}$ quantum numbers, are potential candidates for the radial excited states. In PDG, the $a_{1}(1640)$ was unexpectedly assigned as $n\bar{n}$ member of the $2^{3}P_{1}$ nonet. In past few years, the mass and decay of $n\bar{n}$ member of the $2^{3}P_{1}$ nonet has been analyzed with different models \cite{Godfrey:1985xj,Ebert:2009ub,Pang:2018gcn}. The mass of $a_{1}(1640)$ is obviously less than the theoretical results. The assignment is essential for further testing in future experiments. On the other hand, there is no suitable candidate for the $n\bar{s}$ and $s\bar{s}$ members of the $2^{3}P_{1}$ nonet. In this work, considering the present research situation, we will systematically analyze the masses and decays of the $2^{3}P_{1}$ nonet.

The organization of this paper is as follows. In Section 2, we will provide a concise discussion of the nonrelativistic constituent quark model, Regge phenomenology, and the $^{3}P_{0}$ model. The numerical results of the $2^{3}P_{1}$ meson nonet are presented in Section 3, and Section 4 provides a description of the results.

\section{Theoretical models}
\subsection{Nonrelativistic constituent quark model}
In the constituent quark model, mesons are characterized as bound states of quark and antiquark via a phenomenological potential. It is conventional in the context of the nonrelativistic constituent quark model to assume that the $q\bar{q}$ wave functions is a solution of a nonrelativistic Schr$\ddot{o}$dinger equation with the Breit-Fermi Hamitonian $H$
\cite{Shlyapnikov:2000iw,Lucha:1991vn,Flamm1982}, which can be expressed as
$$
H\psi_{n}(r)=(m_{q}+m_{\bar{q}}+\frac{m_{q}+m_{\bar{q}}}{2m_{q}m_{\bar{q}}}\vec{p^{2}}-(\frac{m_{q}^{3}+m_{\bar{q}}^{3}}{8m_{q}^{3}m_{\bar{q}}^{3}})\vec{p^{4}}
$$
\begin{equation}
\label{Eq1}
+V_{v}(r)+V_{s}(r)+H_{LS+SS+T})\psi_{n}(r)
\end{equation}
with
$$H_{LS+SS+T}=H_{LS}+H_{SS}+H_{T}$$

where $m_{q}$ and $m_{\bar{q}}$ are the masses of constituent quarks, $V_{v}(r)$ and $V_{s}(r)$ are the vector and scalar contributions to the confining potential, and $H_{LS}$, $H_{SS}$ and $H_{T}$ represent the spin-spin, spin-orbit, and tensor terms, respectively.

The following connection may be found in the phenomenological form of the matrix element of the Breit-Fermi Hamitonian when it is applied to P-wave mesons\cite{Li:2006we,Lucha:1991vn,Flamm1982}:

$$
M_{q\bar{q}}=m_{q}+m_{\bar{q}}+a_{1}+b_{1}(\frac{1}{m_{q}}+\frac{1}{m_{\bar{q}}})+c_{1}(\frac{1}{m_{q}^{2}}+\frac{1}{m_{\bar{q}}^{2}})
+\frac{d_{1}}{m_{q}m_{\bar{q}}}$$

$$+g_{1}
[\frac{({m_{q}+m_{\bar{q}}})^{2}+2m_{q}m_{\bar{q}}}{4m_{q}^{2}m_{\bar{q}}^{2}}\langle
L\cdot S\rangle-
\frac{m_{q}^{2}-m_{\bar{q}}^{2}}{4m_{q}^{2}m_{\bar{q}}^{2}}\langle L
(S_{q}-S_{\bar{q}})\rangle]
$$

\begin{equation}
\label{Eq2}
+e_{1}\frac{\langle
S_{q}S_{\bar{q}}\rangle}{m_{q}m_{\bar{q}}}+f_{1}(\frac{1}{m_{q}^{3}}+\frac{1}{m_{\bar{q}}^{3}})+h_{1}\frac{a(j,1)}{m_{q}m_{\bar{q}}}
\end{equation}

with $$ a(j,1)=\frac{4}{5}\langle
S^{2}L^{2}-\frac{3}{2}(LS)-3(LS)^{2}\rangle
$$
$$
\langle LS\rangle=\frac{1}{2}[j(j+1)-3S-2]
$$
where $a_{1}$, $b_{1}$, $c_{1}$, $d_{1}$, $e_{1}$, $f_{1}$, $g_{1}$, and $h_{1}$
are fixed values. In what follows, we employ the Breit-Fermi Hamitonian on the $1^{3}P_{1}$ and $1^{3}P_{2}$ meson multiplets.

$$
M_{q\bar{q}}(1^{3}P_{2})=m_{q}+m_{\bar{q}}+a_{1}+b_{1}(\frac{1}{m_{q}}+\frac{1}{m_{\bar{q}}})+c_{1}(\frac{1}{m_{q}^{2}}+\frac{1}{m_{\bar{q}}^{2}})
+\frac{d_{1}}{m_{q}m_{\bar{q}}}$$

\begin{equation}+g_{1}
\label{Eq3}
\frac{({m_{q}+m_{\bar{q}}})^{2}+2m_{q}m_{\bar{q}}}{4m_{q}^{2}m_{\bar{q}}^{2}}
+\frac{e_{1}}{4m_{q}m_{\bar{q}}}+f_{1}(\frac{1}{m_{q}^{3}}+\frac{1}{m_{\bar{q}}^{3}})-\frac{2h_{1}}{5m_{q}m_{\bar{q}}}
\end{equation}

$$
M_{q\bar{q}}(1^{3}P_{1})=m_{q}+m_{\bar{q}}+a_{1}+b_{1}(\frac{1}{m_{q}}+\frac{1}{m_{\bar{q}}})+c_{1}(\frac{1}{m_{q}^{2}}+\frac{1}{m_{\bar{q}}^{2}})
+\frac{d_{1}}{m_{q}m_{\bar{q}}}$$

\begin{equation}-g_{1}
\label{Eq4}
\frac{({m_{q}+m_{\bar{q}}})^{2}+2m_{q}m_{\bar{q}}}{4m_{q}^{2}m_{\bar{q}}^{2}}
+\frac{e_{1}}{4m_{q}m_{\bar{q}}}+f_{1}(\frac{1}{m_{q}^{3}}+\frac{1}{m_{\bar{q}}^{3}})+\frac{2h_{1}}{m_{q}m_{\bar{q}}}
\end{equation}

From relations (3) and (4), the following relations are derived:

\begin{equation}
\label{Eq5}
\frac{M_{s\bar{s}}(1^{3}P_{2})-M_{s\bar{s}}(1^{3}P_{1})}{M_{n\bar{n}}(1^{3}P_{2})-M_{n\bar{n}}(1^{3}P_{1})}
=
\left(\frac{m_{n}}{m_{s}}\right)^{2}
\end{equation}
where $m_{s}$ and $m_{n}$(n stands for non-strange u- and d-quarks) are masses of the constituent quarks.

\subsection{Regge phenomenology }

In this section, we will review the Regge phenomenology theory. In the 1960s, the Regge theory was developed, which connects the high-energy behavior of the scattering amplitude with singularities in the complex angular momentum plane of the partial wave amplitudes \cite{Regge:1959mz}. Recent years have seen a resurgence in interest in the Regge theory due to the fact that it may be applied to the prediction of meson masses as well as the determination of the quantum numbers of newly detected states in experiments \cite{Li:2004gu,Li:2007px,Guo:2008he,Anisovich:2000kxa,Masjuan:2012gc,Feng:2022hpj,Feng:2022jtg,Chen:2021kfw,Chen:2022flh}. According to Regge theory, mesons have poles that shift in the plane of complex angular momentum as a function of their energy. Regge trajectories of hadrons are often shown on the $(J, M^{2})$ plane, and these plots are known as Chew-Frautschi plots (where $J$ and $M$ are the total spins and masses of the hadrons, respectively).
According to the Chew-Frautschi conjecture, the poles fall onto linear trajectories in the $(J, M^{2})$ plane,

\begin{equation}
\label{Eq6}
J=\alpha_{n\bar{n}N}(0)+\alpha'_{n\bar{n}N}M^2_{n\bar{n}N},
\end{equation}
\begin{equation}
\label{Eq7}
J=\alpha_{n\bar{s}N}(0)+\alpha'_{n\bar{s}N}M^2_{n\bar{s}N},
\end{equation}
\begin{equation}
\label{Eq8}
J=\alpha_{s\bar{s}N}(0)+\alpha'_{s\bar{s}N}M^2_{s\bar{s}N},
\end{equation}
where $N$ is the radial quantum number. The slope and intercept of the Regge trajectory are represented by $\alpha$ and $\alpha'$, respectively. In this work, the intercept and slope can be expressed as
\begin{equation}
\label{Eq9}
\alpha_{n\bar{n}N}(0)+\alpha_{s\bar{s}N}(0)=2\alpha_{n\bar{s}N}(0),
\end{equation}
\begin{equation}
\label{Eq10}
\frac{1}{\alpha'_{n\bar{n}N}}+\frac{1}{\alpha'_{s\bar{s}N}}=\frac{2}{\alpha'_{n\bar{s}N}}.
\end{equation}

The dual-resonance model derived the intercept relation \cite{Berezinsky:1969erk}, which is fulfilled in two-dimensional QCD \cite{Brower:1977as}, the dual-analytic model \cite{Kobylinsky:1978db}, and the quark bremsstrahlung model \cite{Dixit:1979mz}. By employing topological expansion and the $q\bar{q}$-string picture of hadrons, we were able to derive the slope relation (17) \cite{Kaidalov:1980bq}.

From relations (\ref{Eq6})-(\ref{Eq10}), one has
\begin{equation}
\label{Eq11}
M^{2}_{n\bar{n}N}\alpha'_{n\bar{n}N}+M^{2}_{s\bar{s}N}\alpha'_{s\bar{s}N}=2M^{2}_{n\bar{s}N}\alpha'_{n\bar{s}N}.
\end{equation}

Apart from the ground meson, the radial excitations can be estimated in the framework of Regge phenomenology.
Assuming that the ground and the radial excitation have the same slopes\cite{Anisovich:2000kxa}, we derive the following from the relations (\ref{Eq6}), (\ref{Eq7}), and (\ref{Eq8}):
\begin{equation}
\label{Eq12}
M^{2}_{n\overline{n}N}\alpha'_{n\overline{n}1}-M^{2}_{n\bar{n}1}\alpha'_{n\overline{n}1}=\alpha_{n\bar{n}1}(0)-\alpha_{n\bar{n}N}(0),
\end{equation}
\begin{equation}
\label{Eq13}
M^{2}_{s\overline{s}N}\alpha'_{s\overline{s}1}-M^{2}_{s\bar{s}1}\alpha'_{s\overline{s}1}=\alpha_{s\bar{s}1}(0)-\alpha_{s\bar{s}N}(0),
\end{equation}
\begin{equation}
\label{Eq14}
M^{2}_{n\overline{s}N}\alpha'_{n\overline{s}1}-M^{2}_{n\bar{s}1}\alpha'_{n\overline{s}1}=\alpha_{n\bar{s}1}(0)-\alpha_{n\bar{s}N}(0).
\end{equation}

According to Refs. \cite{Filipponi:1997hb,Filipponi:1997vf}, the values of $\alpha_{n\bar{n}1}(0)-\alpha_{n\bar{n}N}(0)$, $\alpha_{n\bar{s}1}(0)-\alpha_{n\bar{s}N}(0)$ and $\alpha_{s\bar{s}1}(0)-\alpha_{s\bar{s}N}(0)$ rely on the masses of the component quarks via the combination $m_{i} + m_{j}$ ($m_{i}$ and $m_{j}$ are the constituent masses of quark and antiquark). In this instance, a factor $f_{i\overline{j}}(m_{i}+m_{j})$ is introduced into relations (\ref{Eq12})-(\ref{Eq14}) \cite{Li:2007px,Liu:2010zzd}. These relations are then stated as

\begin{equation}
\label{Eq15}
M^{2}_{n\bar{n}N}=M^{2}_{n\bar{n}1}+\frac{(N-1)}{\alpha'_{n\bar{n}}}(1+f_{n\bar{n}}(m_{n}+m_{n})),
\end{equation}

\begin{equation}
\label{Eq16}
M^{2}_{n\bar{s}N}=M^{2}_{n\bar{s}1}+\frac{(N-1)}{\alpha'_{n\bar{s}}}(1+f_{n\bar{s}}(m_{n}+m_{s})),
\end{equation}

\begin{equation}
\label{Eq17}
M^{2}_{s\bar{s}N}=M^{2}_{s\bar{s}1}+\frac{(N-1)}{\alpha'_{s\bar{s}}}(1+f_{s\bar{s}}(m_{s}+m_{s})).
\end{equation}

\subsection{$^{3}P_{0}$ model}

Micu proposed the $^{3}P_{0}$ model, and Le Yaouanc refined it \cite{Micu:1968mk,LeYaouanc:1972vsx,LeYaouanc:1974cvx,LeYaouanc:1977fsz}; it is now commonly used to determine the OZI allowed decay processes \cite{Lu:2016bbk,Li:2010vx,Roberts:1992esl,Barnes:1996ff,Ackleh:1996yt,Close:2005se,Barnes:2005pb,Zhang:2006yj,Lu:2014zua,Feng:2022hwq}. In the $^{3}P_{0}$ model, meson decay is caused by the regrouping of the initial meson's $q\bar{q}$  and another $q\bar{q}$ pair produced from vacuum using the quantum numbers $J^{PC} = 0^{++}$.

The transition operator $T$ of the decay $A \rightarrow BC $ in the $^{3}P_{0}$
model is denoted by

$$T=-3 \gamma \sum_m\langle 1 m 1-m \mid 00\rangle \int d^3 \boldsymbol{p}_3 d^3 \boldsymbol{p}_4 \delta^3\left(\boldsymbol{p}_3+\boldsymbol{p}_4\right)$$

$$
\times \mathcal{Y}_1^m\left(\frac{\boldsymbol{p}_3-\boldsymbol{p}_4}{2}\right) \chi_{1,-m}^{34} \phi_0^{34} \omega_0^{34} b_3^{\dagger}\left(\boldsymbol{p}_3\right) d_4^{\dagger}\left(\boldsymbol{p}_4\right)
$$
where $\boldsymbol{p}_3$ and $\boldsymbol{p}_4$ are the momentum of the created quark (antiquark). The dimensionless parameter $\gamma$ represents the strength of the quark-antiquark pair created from
the vacuum. $\chi_{1,-m}^{34}, \phi_0^{34}$, and $\omega_0^{34}$ are spin, flavor, and
color wave functions of the created quark-antiquark pair,
respectively. The partial wave amplitude $\mathcal{M}^{L S}(\boldsymbol{P})$ for the decay $A \rightarrow B+C$ may be written as\cite{Jacob:1959at}

$$
\mathcal{M}^{L S}(\boldsymbol{P})=\sum_{\substack{M_{J_B, M}, M_{J_C} \\ M_S M_L}}\left\langle L M_L S M_S \mid J_A M_{J_A}\right\rangle\left\langle J_B M_{J_B} J_C M_{J_C} \mid S M_S\right\rangle
$$

$$
\int d \Omega Y_{L M_L}^* \mathcal{M}^{M_{J_A} M_{J_B} M_{J_C}(\boldsymbol{P})}
$$

With the transition operator $T$, the helicity amplitude $\mathcal{M}^{M_{J_A} M_{J_B} M_{J_C}}(\boldsymbol{P})$ can be written as
$$
\langle B C|T| A\rangle=\delta^3\left(\boldsymbol{P}_A-\boldsymbol{P}_B-\boldsymbol{P}_C\right) \mathcal{M}^{M_{J_A} M_{J_B} M_{J_C}}(\boldsymbol{P})
$$

Strong decay amplitudes and partial widths are detailed in Refs. \cite{Ackleh:1996yt,Barnes:1996ff}. For the process $A\rightarrow BC$, the partial width is expressed as
\begin{equation}
\label{Eq18}
\Gamma_{A \rightarrow B C}=2 \pi \frac{P E_{B} E_{C}}{M_{A}} \sum_{L S}\left( M_{LS}\right)^{2}
\end{equation}
with
$$
M_{LS}=\frac{\gamma}{\pi^{1 / 4} \beta^{1 / 2}} \xi_{L S}(\frac{P}{\beta}) e^{-P^{2} / 12 \beta^{2}}
$$

$$
P=\frac{\left[\left(M_{A}^{2}-\left(M_{B}+M_{C}\right)^{2}\right)\left(M_{A}^{2}-\left(M_{B}-M_{C}\right)^{2}\right)\right]^{1 / 2}}{2 M_{A}}
$$

$$
E_{B}=\frac{M_{A}^{2}-M_{C}^{2}+M_{B}^{2}}{2 M_{A}}
$$

$$
E_{B}=\frac{M_{A}^{2}+M_{C}^{2}-M_{B}^{2}}{2 M_{A}}
$$

where $P$ is the decay momentum, $E_{B}$ and $E_{C}$ are the energies of meson $B$ and $C$, $M_{A}$, $M_{B}$ are the masses meson $A$ and $B$. The decay amplitude$ M_{LS}$ is proportional to the polynomial $\xi_{L S}(\frac{P}{\beta})$, which is related to decay channels and can be obtained in Refs. \cite{Ackleh:1996yt,Barnes:1996ff}. In this work, we take $\beta=0.4GeV$ and $\gamma=0.4$ as input, which is used in Refs.\cite{Ackleh:1996yt,Barnes:1996ff}.

\section{NUMERICAL RESULTS}

Applying Eqs. (5), (11), (15)-(17) for the meson states $1^{3}P_{2}$ and $2^{3}P_{1}$ states, and incorporating the respective meson and constituent quark masses, we obtain the masses of $M_{n\bar{n}}$, $M_{n\bar{s}}$ and $M_{s\bar{s}}$ members of $2^{3}P_{1}$ states. Our predictions and those given by other references are listed in Table 3. In this work, the parameters used as input are taken from our previous work ( Table 2) \cite{Liu:2010zzd}.

\begin{center}
\indent\\ \footnotesize Table 2. The slopes (GeV$^{-2}$) and the parameters $f_{nn}$ ,$f_{ns}$ and $f_{ss}$ (GeV$^{-1}$) of relations (19), (20), and (21).
\label{Tab:t2}
\begin{tabular}{lllllll}
\hline
\hline
parameters     &$\alpha'_{n\bar{n}}$ & $\alpha'_{n\bar{s}}$ & $\alpha'_{s\bar{s}}$  &$f_{nn}$ & $f_{ns}$ & $f_{ss}$
\\
\hline
value         &  $0.7218$   &$0.6613$    & $0.76902$     & $0.3556$ & $0.1376$  & $0.2219$
\\
\hline

\end{tabular}

\end{center}

\begin{center}
\indent\\ \footnotesize Table 3.  Mass spectrum of the $1^{3}P_{1}$ and $2^{3}P_{1}$ meson nonets (in units of MeV). The masses used as input for our calculation
are shown in boldface.
\label{Tab:t3}
\begin{tabular}{llllllllllll}

\hline
\hline
  state                     &           &$M_{n\bar{n}}$ &           &$M_{n\bar{s}}$&                &$M_{s\bar{s}}$ &
\\
\hline
 $Refs.$                      &   $1^{3}P_{1}$           &   $2^{3}P_{1}$        & $1^{3}P_{1}$        & $2^{3}P_{1}$    &   $1^{3}P_{1}$    & $2^{3}P_{1}$
\\
 $Present$                   &   $\mathbf{{1230}\pm40}$  &   1784.4              &  1366.2             & 1885.1     &   1498.2         & 1986.0
\\
 Ref. \cite{Godfrey:1985xj}  &     1240                  &   1820                &  1380               & 1930       &   1480           &  2030
\\
 Ref. \cite{Ebert:2009ub}   &    1254                   &   1742                &  1412              &  1893        &   1464          &   2016
\\
 Ref. \cite{Xiao:2019qhl}   &                           &                       &                     &            &   1480            &  2027
\\
 Ref. \cite{Li:2020xzs}     &                           &                       &                     &            &   1492            &   2027
\\

\hline
\hline
\end{tabular}

\end{center}


Employing the $^{3}P_{0}$ model, the decays of $2^{3}P_{1}(n\bar{n})$, $2^{3}P_{1}(n\bar{s})$  and $2^{3}P_{1}(s\bar{s})$ are investigated, and the results are listed in Tab.? and Tab.?. $KK_{1}(1270)$ depends on the mixing of $K_{1A}$ and $K_{1B}$ states, $ K_{1}(1270)=K_{1}\left({1}^{1}P_{1}\right) \cos \theta_{K}+K_{1}\left({1}^{3} P_{1}\right) \sin \theta_{K}$, the $\theta_{K}$ denotes the mixing angle. The mixing angle is investigated in the Refs.\cite{Divotgey:2013jba, Blundell:1995au,Pang:2017dlw}. In the present work, we take $\theta_{K}=45^{o}$ as input parameters \cite{Blundell:1995au,Pang:2017dlw}.

\begin{center}
\indent\\ \footnotesize Table 4. Strong decay properties for the $2^{3}P_{1}(n\bar{s})$ state (in units of MeV). The masses $M_{2^{3}P_{1}(n\bar{n})}=1784.4$ MeV, $M_{2^{3}P_{1}(s\bar{s})}=1986$ MeV  and the masses of all the final states are taken from PDG.
\label{Tab:t2}
\label{Tab:t2}
\begin{tabular}{llllllllllll}

\hline
\hline
  Decay mode                                   &Present work & Decay mode                                 & Present work
\\
\hline
 $2^{3}P_{1}(n\bar{n})\rightarrow \rho \pi $        & 76.5     & $2^{3}P_{1}(s\bar{s})\rightarrow K K^{*}      $          & 78
\\
 $2^{3}P_{1}(n\bar{n})\rightarrow \omega \rho$      & 42       & $2^{3}P_{1}(s\bar{s})\rightarrow K^{*}K^{*}   $          & 48
\\
$2^{3}P_{1}(n\bar{n})\rightarrow \rho(1465) \pi$    & 58       & $2^{3}P_{1}(s\bar{s})\rightarrow K K_{1}(1270)$          & 72.7
\\
$2^{3}P_{1}(n\bar{n})\rightarrow b_{1}(1230)\pi$   & 62.3      & $2^{3}P_{1}(s\bar{s})\rightarrow K K_{1}(1400) $         & 25.5
\\
 $2^{3}P_{1}(n\bar{n})\rightarrow f{1}(1285)\pi $  & 27.8      & $2^{3}P_{1}(s\bar{s})\rightarrow K K_{0}^{*}(1430) $      & 1.9
\\
$2^{3}P_{1}(n\bar{n})\rightarrow f{2}(1275)\pi$    & 61.5      & $2^{3}P_{1}(s\bar{s})\rightarrow K K_{2}^{*}(1430) $      & 31.6
\\
$2^{3}P_{1}(n\bar{n})\rightarrow K^{*}K $          & 5.5       & $2^{3}P_{1}(s\bar{s})\rightarrow K K_{2}^{*}(1414)$       & 94

\\
\hline
\hline
\end{tabular}

\end{center}


\begin{center}
\indent\\ \footnotesize Table 5. Strong decay properties for the $2^{3}P_{1}(n\bar{s})$ state (in units of MeV). The mass $M_{2^{3}P_{1}(n\bar{s})}=1885.1$ MeV and the masses of all the final states are taken from PDG.
\label{Tab:t2}
\label{Tab:t2}
\begin{tabular}{llllllllllll}

\hline
\hline
  Decay mode                                   &Present work & Decay mode                                 & Present work
\\
\hline
 $2^{3}P_{1}(n\bar{s})\rightarrow \rho K  $        & 27        & $2^{3}P_{1}(n\bar{s})\rightarrow b_{1 }K       $          & 16.1
\\
 $2^{3}P_{1}(n\bar{s})\rightarrow \omega K$        & 8.8       & $2^{3}P_{1}(n\bar{s})\rightarrow h_{1 }K    $          & 12.3
\\
$2^{3}P_{1}(n\bar{s})\rightarrow \phi K   $        & 20.7      & $2^{3}P_{1}(n\bar{s})\rightarrow a_{1 }K  $          & 19.1
\\
$2^{3}P_{1}(n\bar{s})\rightarrow \pi K^{*} $        & 30.0      & $2^{3}P_{1}(n\bar{s})\rightarrow f_{1 }K   $         & 8.2
\\
$2^{3}P_{1}(n\bar{s})\rightarrow \eta K^{*} $        & 34.8     & $2^{3}P_{1}(n\bar{s})\rightarrow f_{2 }K   $         & 13.9
\\
$2^{3}P_{1}(n\bar{s})\rightarrow \rho K^{*} $        & 35.5     & $2^{3}P_{1}(n\bar{s})\rightarrow \pi K_{1}(1270) $         & 31.0
\\
$2^{3}P_{1}(n\bar{s})\rightarrow \omega K^{*} $        & 10.9     & $2^{3}P_{1}(n\bar{s})\rightarrow \pi K_{1}(1400) $         & 20
\\
$2^{3}P_{1}(n\bar{s})\rightarrow K K_{2}^{*}(1414)$       & 23.2     & $2^{3}P_{1}(n\bar{s})\rightarrow K K_{0}^{*}(1430) $      & 25
\\
$2^{3}P_{1}(n\bar{s})\rightarrow K K_{2}^{*}(1430) $      & 40.1    &

\\
\hline
\hline
\end{tabular}

\end{center}


\section{Conclusion}
In the present work, we examine the mass spectrum of the $2^{3}P_{1}$ meson nonet by merging the non-relativistic component quark model with Regge phenomenology. The mass of $n\bar{n}$ is determined to be $1784.4$ MeV, which is consistent with different theoretical predictions \cite{Godfrey:1985xj,Ebert:2009ub}. Considering the fact that the values is about $120$ MeV higher than the measured mass of $a_{1}(1640)$, we suggest that
the assignment for the $a_{1}(1640)$ as $n\bar{n}$ member of the $2^{3}P_{1}$ meson nonet need further testing in the experiment. Moreover, due to poor information on $n\bar{s}$ and $s\bar{s}$ member of $2^{3}P_{1}$ meson nonet, the investigation of these states has become a fascinating problem. In the present work, $n\bar{s}$ and $s\bar{s}$ members are determined to be 1885.1 MeV and 1996 MeV, respectively. The findings could be able to give a meaningful mass range for the phenomenological investigation. Apart from the mass spectrum, we propose the decays of these states. It is possible that in the near future it will be essential to carry out more screening, both theoretically and experimentally, for probable candidates of radial excited states. The results of our analysis could provide some benefit from this work.


\begin{thebibliography}{99}
{
\bibitem{ParticleDataGroup:2022pth}
R.~L.~Workman \textit{et al.} [Particle Data Group],
PTEP \textbf{2022}, 083C01 (2022)
doi:10.1093/ptep/ptac097


\bibitem{Li:2006we}
D.~M.~Li and Z.~Li,
Eur. Phys. J. A \textbf{28}, 369-373 (2006)
doi:10.1140/epja/i2006-10067-y
[arXiv:hep-ph/0606297 [hep-ph]].

\bibitem{Cheng:2011pb}
H.~Y.~Cheng,
Phys. Lett. B \textbf{707}, 116-120 (2012)
doi:10.1016/j.physletb.2011.12.013
[arXiv:1110.2249 [hep-ph]].

\bibitem{Divotgey:2013jba}
F.~Divotgey, L.~Olbrich and F.~Giacosa,
Eur. Phys. J. A \textbf{49}, 135 (2013)
doi:10.1140/epja/i2013-13135-3
[arXiv:1306.1193 [hep-ph]].

\bibitem{Zhang:2017cbi}
Z.~Q.~Zhang, H.~Guo and S.~Y.~Wang,
Eur. Phys. J. C \textbf{78}, no.3, 219 (2018)
doi:10.1140/epjc/s10052-018-5674-7
[arXiv:1705.00524 [hep-ph]].


\bibitem{Godfrey:1985xj}
S.~Godfrey and N.~Isgur,
Phys. Rev. D \textbf{32}, 189-231 (1985)
doi:10.1103/PhysRevD.32.189


\bibitem{Ebert:2009ub}
D.~Ebert, R.~N.~Faustov and V.~O.~Galkin,
Phys. Rev. D \textbf{79}, 114029 (2009)
doi:10.1103/PhysRevD.79.114029
[arXiv:0903.5183 [hep-ph]].

\bibitem{Pang:2018gcn}
C.~Q.~Pang, Y.~R.~Wang and C.~H.~Wang,
Phys. Rev. D \textbf{99}, no.1, 014022 (2019)

\bibitem{Shlyapnikov:2000iw}
P.~V.~Shlyapnikov,
Phys. Lett. B \textbf{496}, 129-136 (2000)
doi:10.1016/S0370-2693(00)01286-7

\bibitem{Lucha:1991vn}
W.~Lucha, F.~F.~Sch$\ddot{o}$berl and D.~Gromes,
Phys. Rept. \textbf{200}, 127-240 (1991)
doi:10.1016/0370-1573(91)90001-3


\bibitem{Flamm1982}D. Flamm, F. F. Sch$\ddot{o}$berl, Introduction to quark model of elementary particles, Vol. 1, Gordon
and Breach Science Publishers, London, 1982


\bibitem{Regge:1959mz}
T.~Regge,
``Introduction to complex orbital momenta,''
Nuovo Cim. \textbf{14}, 951 (1959).

\bibitem{Li:2004gu}
D.~M.~Li, B.~Ma, Y.~X.~Li, Q.~K.~Yao and H.~Yu,
``Meson spectrum in Regge phenomenology,''
Eur. Phys. J. C \textbf{37}, 323 (2004).

\bibitem{Li:2007px}
D.~M.~Li, B.~Ma and Y.~H.~Liu,
``Understanding masses of $c\bar{s}$ states in Regge phenomenology,''
Eur. Phys. J. C \textbf{51}, 359 (2007).

\bibitem{Guo:2008he}
X.~H.~Guo, K.~W.~Wei and X.~H.~Wu,
``Some mass relations for mesons and baryons in Regge phenomenology,''
Phys. Rev. D \textbf{78}, 056005 (2008).

\bibitem{Anisovich:2000kxa}
A.~V.~Anisovich, V.~V.~Anisovich and A.~V.~Sarantsev,
``Systematics of $q\bar{q}$ states in the (n, $M^{2}$) and (J, $M^{2}$) planes,''
Phys. Rev. D \textbf{62}, 051502 (2000).

\bibitem{Masjuan:2012gc}
P.~Masjuan, E.~Ruiz Arriola and W.~Broniowski,
``Systematics of radial and angular-momentum Regge trajectories of light non-strange $q\bar{q}$ states,''
Phys. Rev. D \textbf{85}, 094006 (2012).

\bibitem{Feng:2022hpj}
X.~C.~Feng, K.~W.~Wei, Jie-Wu and J.~Wu,
Eur. Phys. J. A \textbf{58}, no.11, 233 (2022)
doi:10.1140/epja/s10050-022-00886-5
[arXiv:2211.07083 [hep-ph]].

\bibitem{Feng:2022jtg}
X.~C.~Feng, K.~Wei Wei, J.~Wu, X.~Z.~Zhai and S.~Wang,
Acta Phys. Polon. B \textbf{53}, no.10, 4 (2022)
doi:10.5506/APhysPolB.53.10-A4
[arXiv:2211.03921 [hep-ph]].

\bibitem{Chen:2021kfw}
J.~K.~Chen,
Eur. Phys. J. A \textbf{57}, no.7, 238 (2021)
doi:10.1140/epja/s10050-021-00502-y
[arXiv:2102.07993 [hep-ph]].

\bibitem{Chen:2022flh}
J.~K.~Chen,
Nucl. Phys. B \textbf{983}, 115911 (2022)
doi:10.1016/j.nuclphysb.2022.115911
[arXiv:2203.02981 [hep-ph]].

\bibitem{Berezinsky:1969erk}
V.~S.~Berezinsky and G.~T.~Zatsepin,
Phys. Lett. B \textbf{28}, 423-424 (1969)
doi:10.1016/0370-2693(69)90341-4

\bibitem{Brower:1977as}
R.~C.~Brower, J.~R.~Ellis, M.~G.~Schmidt and J.~H.~Weis,
Nucl. Phys. B \textbf{128}, 175-203 (1977)
doi:10.1016/0550-3213(77)90304-2

\bibitem{Kobylinsky:1978db}
N.~A.~Kobylinsky, E.~S.~Martynov and A.~B.~Prognimak,
Ukr. Fiz. Zh. (Russ. Ed. ) \textbf{24}, 969-974 (1979)
ITF-78-93E.



\bibitem{Dixit:1979mz}
V.~V.~Dixit and L.~A.~P.~Balazs,
Phys. Rev. D \textbf{20}, 816 (1979)
doi:10.1103/PhysRevD.20.816

\bibitem{Kaidalov:1980bq}
A.~B.~Kaidalov,
Z. Phys. C \textbf{12}, 63 (1982)
doi:10.1007/BF01475732



\bibitem{Filipponi:1997hb}
S.~Filipponi, G.~Pancheri and Y.~Srivastava,
Phys. Rev. Lett. \textbf{80}, 1838-1840 (1998)

\bibitem{Filipponi:1997vf}
S.~Filipponi and Y.~Srivastava,
Phys. Rev. D \textbf{58}, 016003 (1998)

\bibitem{Liu:2010zzd}
Y.~H.~Liu, X.~C.~Feng and J.~J.~Zhang,
Eur. Phys. J. A \textbf{43}, 379-388 (2010)
doi:10.1140/epja/i2010-10922-2

\bibitem{Micu:1968mk}
L.~Micu,
Nucl. Phys. B \textbf{10}, 521-526 (1969)
doi:10.1016/0550-3213(69)90039-X

\bibitem{LeYaouanc:1972vsx}
A.~Le Yaouanc, L.~Oliver, O.~Pene and J.~C.~Raynal,
Phys. Rev. D \textbf{8}, 2223-2234 (1973)
doi:10.1103/PhysRevD.8.2223

\bibitem{LeYaouanc:1974cvx}
A.~Le Yaouanc, L.~Oliver, O.~Pene and J.~C.~Raynal,
Phys. Rev. D \textbf{11}, 1272 (1975)
doi:10.1103/PhysRevD.11.1272

\bibitem{LeYaouanc:1977fsz}
A.~Le Yaouanc, L.~Oliver, O.~Pene and J.~C.~Raynal,
Phys. Lett. B \textbf{71}, 397-399 (1977)
doi:10.1016/0370-2693(77)90250-7

\bibitem{Lu:2016bbk}
Q.~F.~Lu, T.~T.~Pan, Y.~Y.~Wang, E.~Wang and D.~M.~Li,
Phys. Rev. D \textbf{94}, no.7, 074012 (2016)
doi:10.1103/PhysRevD.94.074012
[arXiv:1607.02812 [hep-ph]].


\bibitem{Li:2010vx}
D.~M.~Li, P.~F.~Ji and B.~Ma,
Eur. Phys. J. C \textbf{71}, 1582 (2011)
doi:10.1140/epjc/s10052-011-1582-9
[arXiv:1011.1548 [hep-ph]].

\bibitem{Roberts:1992esl}
W.~Roberts and B.~Silvestre-Brac,
Few Body Syst. \textbf{11}, no.4, 171-193 (1992)
doi:10.1007/bf01641821

\bibitem{Ackleh:1996yt}
E.~S.~Ackleh, T.~Barnes and E.~S.~Swanson,
Phys. Rev. D \textbf{54}, 6811-6829 (1996)
doi:10.1103/PhysRevD.54.6811
[arXiv:hep-ph/9604355 [hep-ph]].

\bibitem{Barnes:1996ff}
T.~Barnes, F.~E.~Close, P.~R.~Page and E.~S.~Swanson,
Phys. Rev. D \textbf{55}, 4157-4188 (1997)
doi:10.1103/PhysRevD.55.4157
[arXiv:hep-ph/9609339 [hep-ph]].

\bibitem{Close:2005se}
F.~E.~Close and E.~S.~Swanson,
Phys. Rev. D \textbf{72}, 094004 (2005)
doi:10.1103/PhysRevD.72.094004
[arXiv:hep-ph/0505206 [hep-ph]].

\bibitem{Barnes:2005pb}
T.~Barnes, S.~Godfrey and E.~S.~Swanson,
Phys. Rev. D \textbf{72}, 054026 (2005)
doi:10.1103/PhysRevD.72.054026
[arXiv:hep-ph/0505002 [hep-ph]].

\bibitem{Zhang:2006yj}
B.~Zhang, X.~Liu, W.~Z.~Deng and S.~L.~Zhu,
Eur. Phys. J. C \textbf{50}, 617-628 (2007)
doi:10.1140/epjc/s10052-007-0221-y
[arXiv:hep-ph/0609013 [hep-ph]].

\bibitem{Lu:2014zua}
Q.~F.~L\"u and D.~M.~Li,
Phys. Rev. D \textbf{90}, no.5, 054024 (2014)
doi:10.1103/PhysRevD.90.054024
[arXiv:1407.3092 [hep-ph]].

\bibitem{Feng:2022hwq}
X.~C.~Feng, Z.~Y.~Li, D.~M.~Li, Q.~T.~Song, E.~Wang and W.~C.~Yan,
Phys. Rev. D \textbf{106}, no.7, 076012 (2022)
doi:10.1103/PhysRevD.106.076012
[arXiv:2206.10132 [hep-ph]].

\bibitem{Jacob:1959at}
M.~Jacob and G.~C.~Wick,
Annals Phys. \textbf{7}, 404-428 (1959)
doi:10.1016/0003-4916(59)90051-X

\bibitem{Xiao:2019qhl}
L.~Y.~Xiao, X.~Z.~Weng, X.~H.~Zhong and S.~L.~Zhu,
Chin. Phys. C \textbf{43}, no.11, 113105 (2019)
doi:10.1088/1674-1137/43/11/113105
[arXiv:1904.06616 [hep-ph]].

\bibitem{Li:2020xzs}
Q.~Li, L.~C.~Gui, M.~S.~Liu, Q.~F.~L\"u and X.~H.~Zhong,
Chin. Phys. C \textbf{45}, no.2, 023116 (2021)
doi:10.1088/1674-1137/abcf22
[arXiv:2004.05786 [hep-ph]].



\bibitem{Blundell:1995au}
H.~G.~Blundell, S.~Godfrey and B.~Phelps,
Phys. Rev. D \textbf{53}, 3712-3722 (1996)
doi:10.1103/PhysRevD.53.3712
[arXiv:hep-ph/9510245 [hep-ph]].

\bibitem{Pang:2017dlw}
C.~Q.~Pang, J.~Z.~Wang, X.~Liu and T.~Matsuki,
``A systematic study of mass spectra and strong decay of strange mesons,''
Eur. Phys. J. C \textbf{77}, no.12, 861 (2017).

}

\end{thebibliography}
\end{document}